\documentclass{kapproc} 
\kluwerbib

\begin{document}
\articletitle[Timescale Spectra in High Energy Astrophysics]
{Timescale Spectra\\
 in High Energy Astrophysics}

\author{T.P. Li}
\affil{Department of Physics and Center for Astrophysics, Tsinghua University\\
Institute of High Energy Physics, Chinese Academy of Sciences\\
Beijing, China \footnote{Supported by the Special funds for Major State
Basic Research Projects of China}}
\email{litp@mail.tsinghua.edu.cn}

\begin{abstract}
A technique of timescale analysis performed directly in the time domain
has been developed recently. We have applied the technique to study 
rapid variabilities of hard X-rays from neutron star and black hole
binaries, $\gamma$-ray bursts and terrestrial $\gamma$-ray flashes.
 The results indicate that the time domain method of spectral analysis
is a powerful tool in revealing the underlying physics in high-energy processes
in objects.  
\end{abstract}


\section{Introduction}
The widely used Fourier analysis method in temporal analysis 
is to derive frequency spectra from a time series.
After the Fourier transform of a light curve $x(t_k)$
\begin{equation} 
X(f_j)=\sum_k x(t_k) e^{-i2\pi f_jk\Delta t}, \hspace{5mm} f_j=j/T 
\end{equation}
we can get the power density spectrum
\begin{equation} P_j=|X(f_j)|^2 \end{equation} 
From two light curves, $x_1(t_k)$ and $x_2(t_k)$, observed simultaneously in two energy 
bands at times $t_k$, and their Fourier transforms $X_1(f_j)$ and $X_2(f_j)$, we can 
construct the cross spectrum 
\begin{equation}C(f_j)=X_1^*(f_j)X_2(f_j) \end{equation}
and then the time lag spectrum 
\begin{equation} \Lambda (f_j)=\arg [C(f_j)]/2\pi f_j \end{equation}
and the coherence coefficient spectrum 
\begin{equation} r(f)=\frac{|<C(f)>|}{\sqrt{<|X_1(f)|^2><|X_2(f)|^2>}} \end{equation}

People usually take a Fourier period $1/f$ as a timescale and use the Fourier power
spectrum to describe the distribution of variation amplitude at different timescales,
the time lag spectrum to describe the distribution of the emission time difference between
two energy bands at different timescales and the coherence coefficient spectrum 
to describe the distribution of degree of linear correlation between high and low energy
processes at different timescales. But it is not correct to equate the Fourier period 
with the timescale; a Fourier component with a certain frequency $f$ of a light curve
is not equal to the real process with the timescale $1/f$. For some important 
high-energy emission processes the Fourier spectra distort the timescale distribution
of real physical processes seriously. It is needed to derive timescale spectra from
observed light curves directly in the time domain without through Fourier transforms.   

\section{Power Spectra}
The power density spectrum can be derived directly in the time domain (Li 2001). 
For a counting series $x(k),~k=1,...,N$ obtained from a time history of observed photons
with a time step $\Delta t$,
the definition of variation power is
\begin{equation}
P(\Delta t)=\frac{\mbox{Var}(x)}{(\Delta t)^2}=\frac{\frac{1}{N}\sum_{k=1}^{N}(x(k)-\bar{x})^2}
{(\Delta t)^2}=\frac{1}{N}\sum_{k=1}^{N}(r(k)-\bar{r})^2
\hspace{3mm} \mbox{rms$^2$} 
\end{equation}
where $r=x/
\Delta t$.
The power density $p(\Delta t)$ at a timescale $\Delta t$ can be then derived
\begin{equation}
p(\Delta t)=\frac{\mbox{d} P(\Delta t)}{\mbox{d} \Delta t}\simeq\frac{P(\Delta t_1)-P(\Delta t_2)}{\Delta t_2-\Delta t_1}
 \hspace{7mm} \mbox{rms$^2$/s} \end{equation}
From (6), (7) we can calculate the power density for Poison noise
\begin{equation}
p_{noise}(\Delta t)\simeq\frac{r}{\Delta t_1 \Delta t_2} \hspace{7mm} \mbox{rms$^2$/s} 
\end{equation}
and the signal power density can be defined as
\begin{equation}
p_{signal}(\Delta t)=p(\Delta t)-p_{noise}(\Delta t) \hspace{7mm} \mbox{rms$^2$/s} 
 \end{equation}

\begin{figure}[ht]
\includegraphics{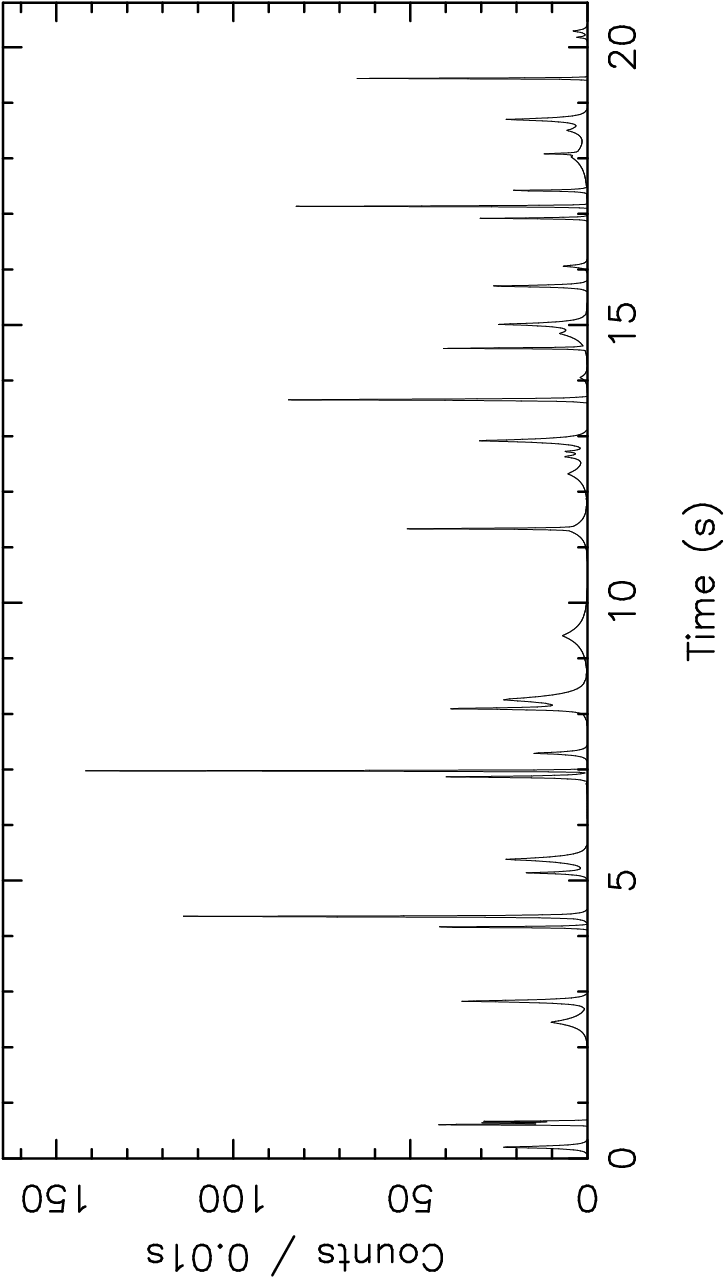}
\includegraphics{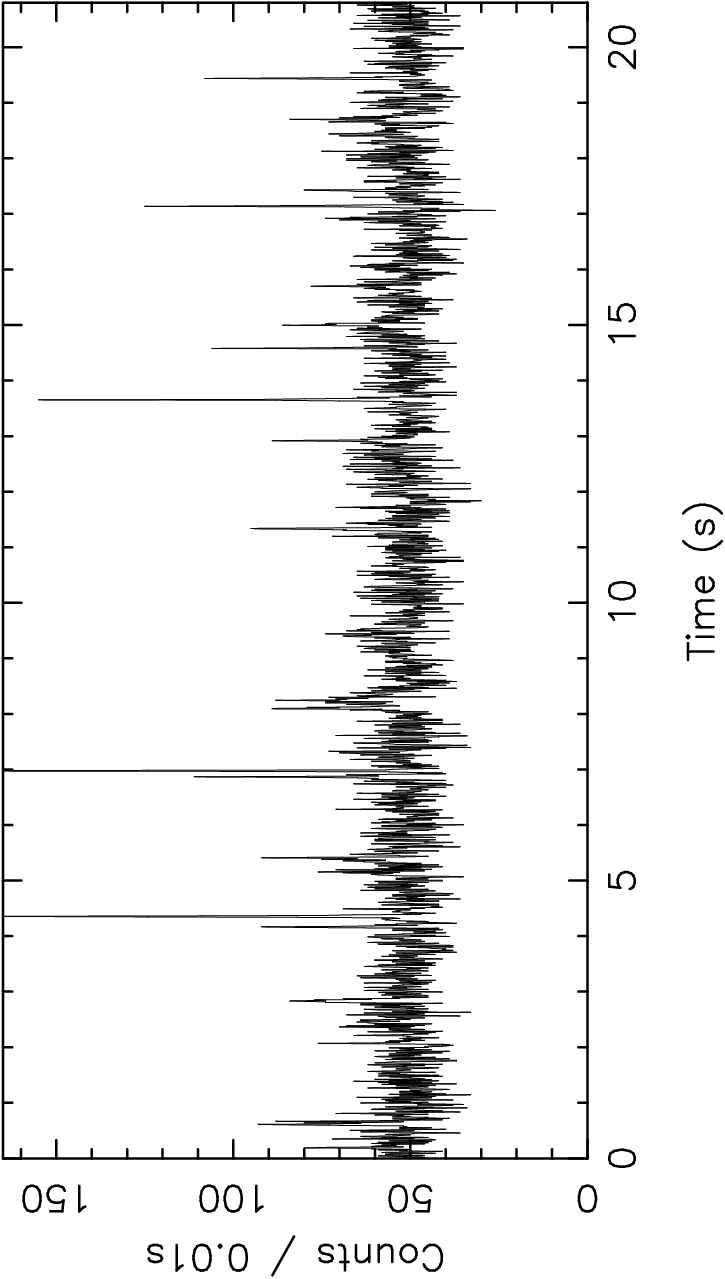}
\includegraphics{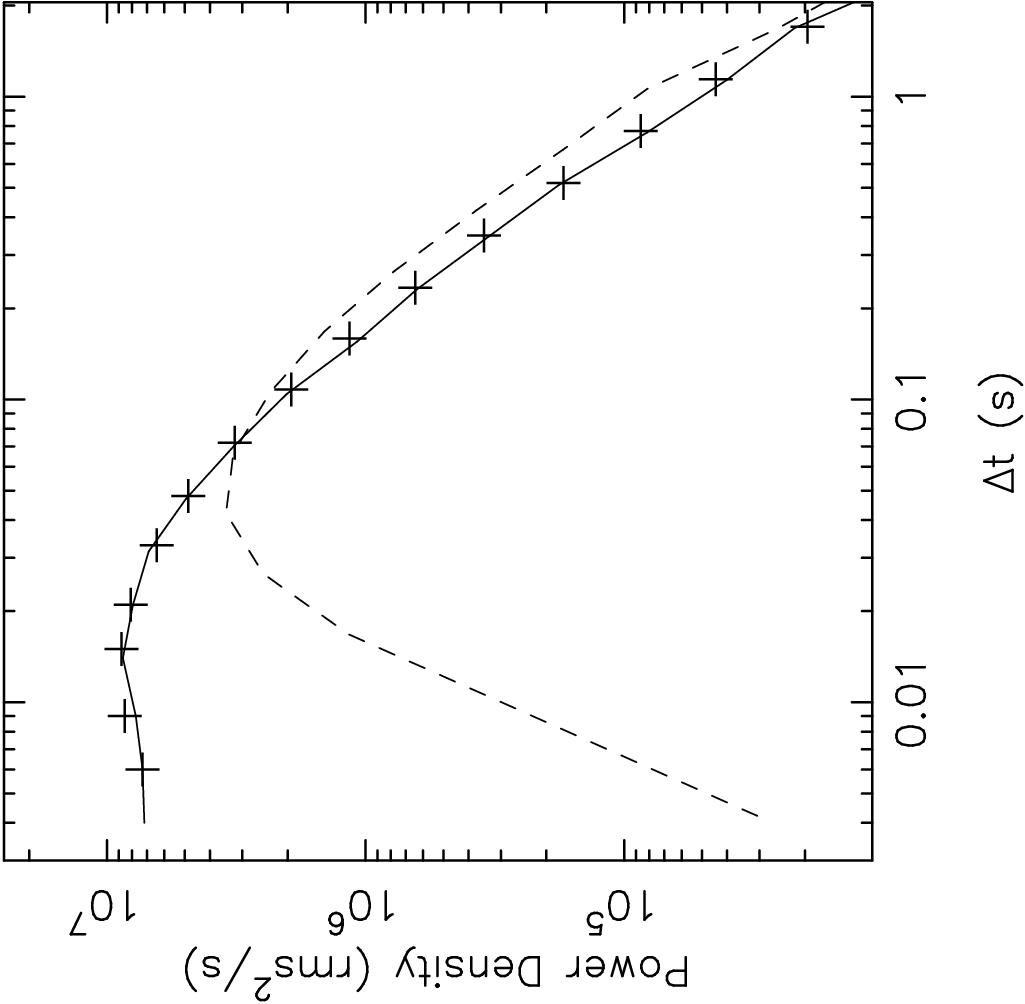}
\vspace{4mm}
\narrowcaption{Distribution of power density vs. time scale of a shot model. 
{\sl Top panel}: the signal, stochastic exponential shots with time constant
$\tau$ 
between 5 ms and 0.2 s. {\sl Middle panel}: simulated data which involved both
 signal and Poisson noises.
{\sl Bottom panel}: signal power densities . {\it Solid line} -- power density
 distribution of time scale expected for the signal. 
{\it Dashed line} -- excess Fourier spectrum from the simulated data.
{\it Plus signs} -- excess power densities 
calculated by the timing technique in the time domain for the simulated data. }
\vspace{3mm}
\end{figure}

In Fig. 1 the bottom panel shows the expected power spectrum (solid line) for 
a signal consisting of stochastic shots (shown in the top panel), 
the timescale spectrum of
power density (crosses) and Fourier spectrum (dashed line) from the simulated 
data (shown in the middle panel) of the stochastic signal. 
As the rise and decay time constant
of a shot is randomly taken from the range between 5 ms and 0.2 s, there should exist
significant variation power in this timescale region. One can see from Fig. 1 that the 
Fourier spectrum significantly underestimates the power densities at the
shorter timescales.

Timescale spectra of power density can help to reveal the nature of physical process
around compact stars (Li \& Muraki 2002).  Figure 2 shows the timescale spectra 
and Fourier spectra of power density for a sample of accreting black holes (left side) 
and neutron stars (right side). The five black hole candidates demonstrate
a significant power excess in the timescale spectra in comparison with their
corresponding Fourier spectra at timescales shorter than $\sim 0.1$ s, but the
two kinds of power spectrum in the studied neutron stars are generally
consistent with each other. Assuming that there exist a stochastic process with 
a characteristic time of $\sim 0.1$ s in the black hole systems and that 
a significant variability of the accreting neutron stars comes from stochastic 
processes with characteristic time constants much shorter than $\sim 1$ ms can explain 
the observed power spectra for X-ray binaries. 

\begin{figure}[ht]
\includegraphics{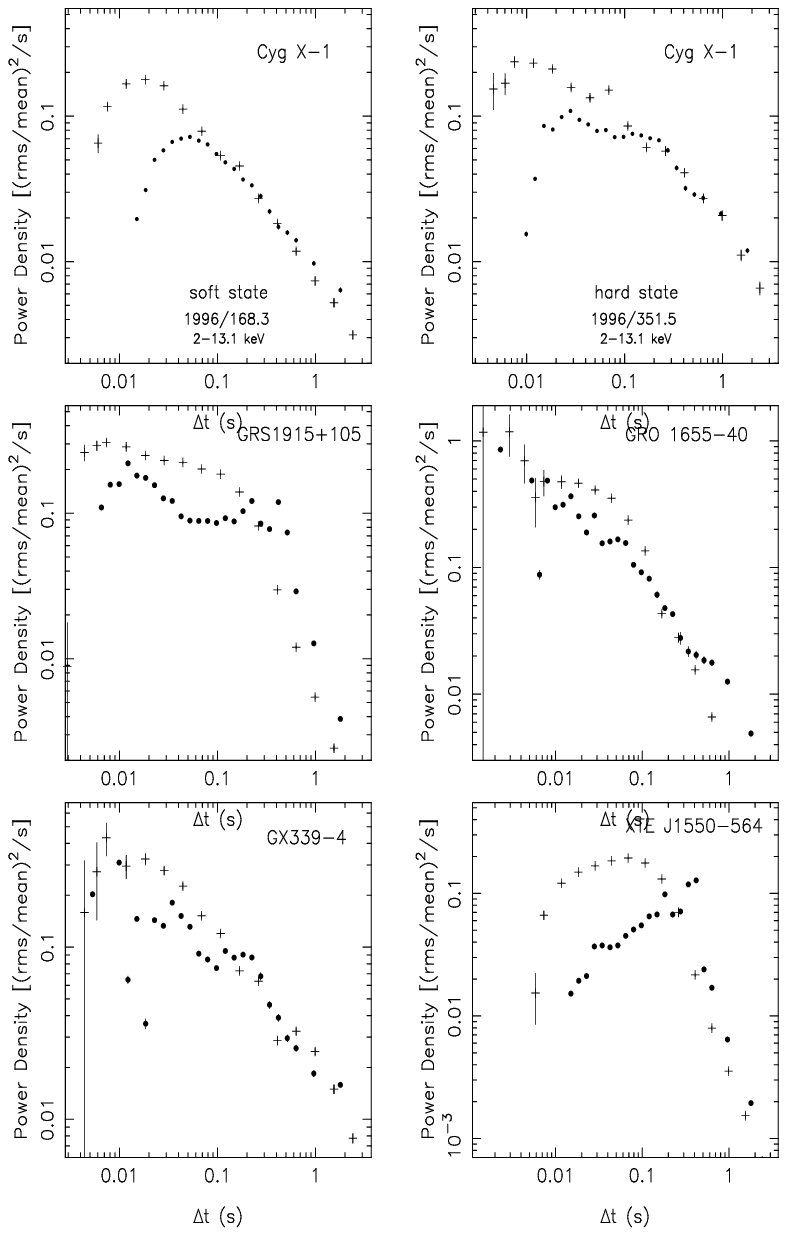}
\includegraphics{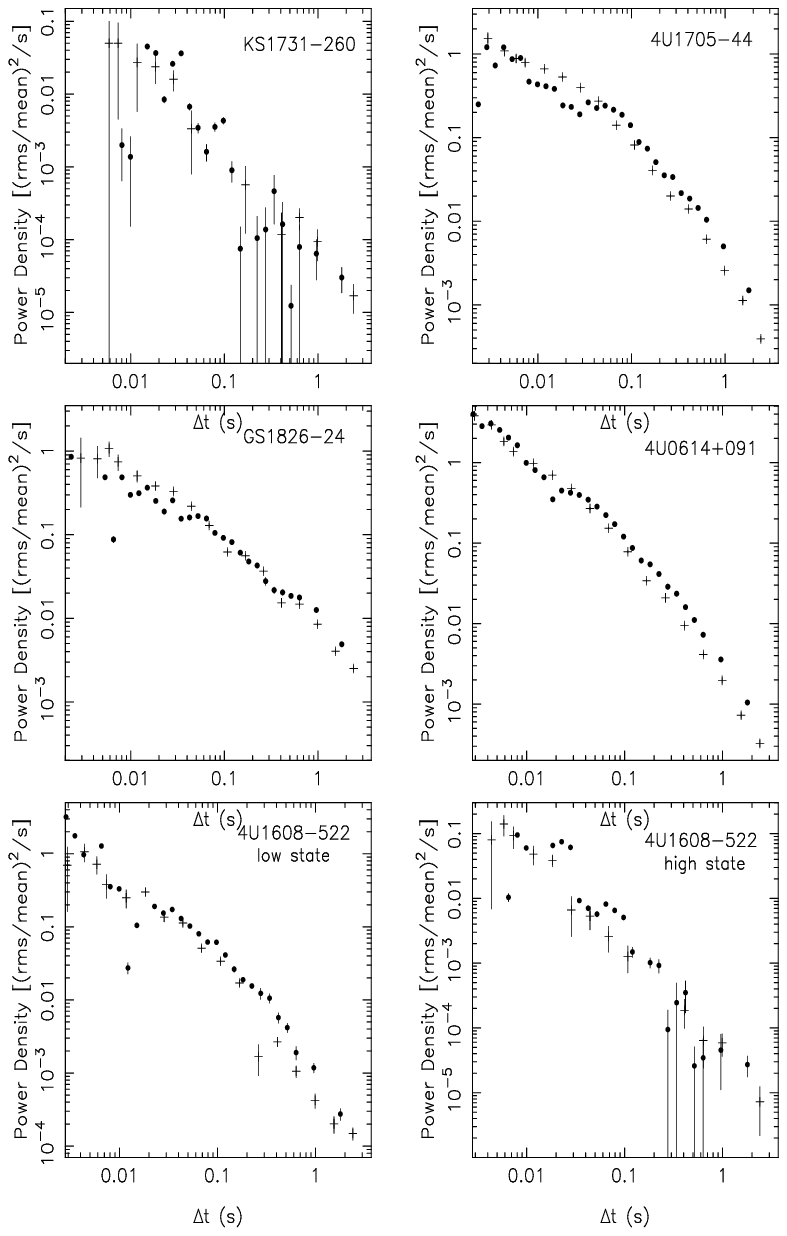}
\vspace{95mm}
\caption{Power density vs. timescale of black hole binaries (left side)
and neutron star binaries (right side). $Crosses$: Timescale spectra.
$Dots$: Fourier spectra}
\end{figure}

\section{Time Lag Spectra}
Measuring the relative time delay between photons in two energy bands can 
give useful clue to understanding the emission mechanism and emitting region
(e.g., Kazanas, Hua \& Titarchuk 1997; Hua, Kazanas \& Titarchuk 1997; 
Nowak et al. 1999).
With the cross-correlation analysis one can calculate a time lag $\Lambda$
from two light curves. Traditional cross-correlation
method fails to calculate any time lag shorter than the time step $\Delta t$. 
In studying a complex process only a time lag $\Lambda$ 
is not enough, we need to know time lags at different timescales,
i.e. the timescale spectrum $\Lambda(\Delta t)$.  Using Fourier analysis
we can get a time lag spectrum. But as the Fourier technique is powerless
for detecting the variation power of a stochastic process at high frequencies
(short timescales) range, the Fourier analysis method is also powerless
for detecting the time lags at high frequencies (short timescales). 

The timescale spectrum of time lag can be derived with a modified 
cross-correlation technique (Li, Feng \& Chen 1999; Li 2001).
For two light curves $x_1(k),~x_2(k)$ with a time step $\Delta t$ the modified 
cross-correlation function at $\tau$ is defined as
\begin{equation} \mbox{C}(\tau; \Delta t)=\sum_k v_1(k\Delta t)v_2(k\Delta t+\tau)/\sigma(v_1)\sigma(v_2) 
\end{equation}
where $v(t)=x(t)-\bar{x}$, $\sigma^2(v)=\sum v^2$, $x(t)$ is the counts in the time interval
$(t, t+\Delta t)$. 
Find the time lag $\Lambda$ letting $C(\tau=\Lambda; \Delta t)=\max$ for different timescale $\Delta t$, 
we can derive the timescale spectrum of time lag, $\Lambda(\Delta)$.

For comparing the two kinds of spectrum for time lags, timescale spectrum and
Fourier spectrum, we construct two simulated time event series which
consist of two kinds of random shot component : rapid components with typical
timescale 1 ms and time lag 5 ms and slow component with timescale 0.01 s and
time lag 0.02 s. Figure 3 shows that the timescale spectrum can represent
the time delay distribution in the real process, but the Fourier analysis
is powerless to detecting the time lags in the short timescale range.
 
\begin{figure}[ht]
\vspace{5mm}
\includegraphics{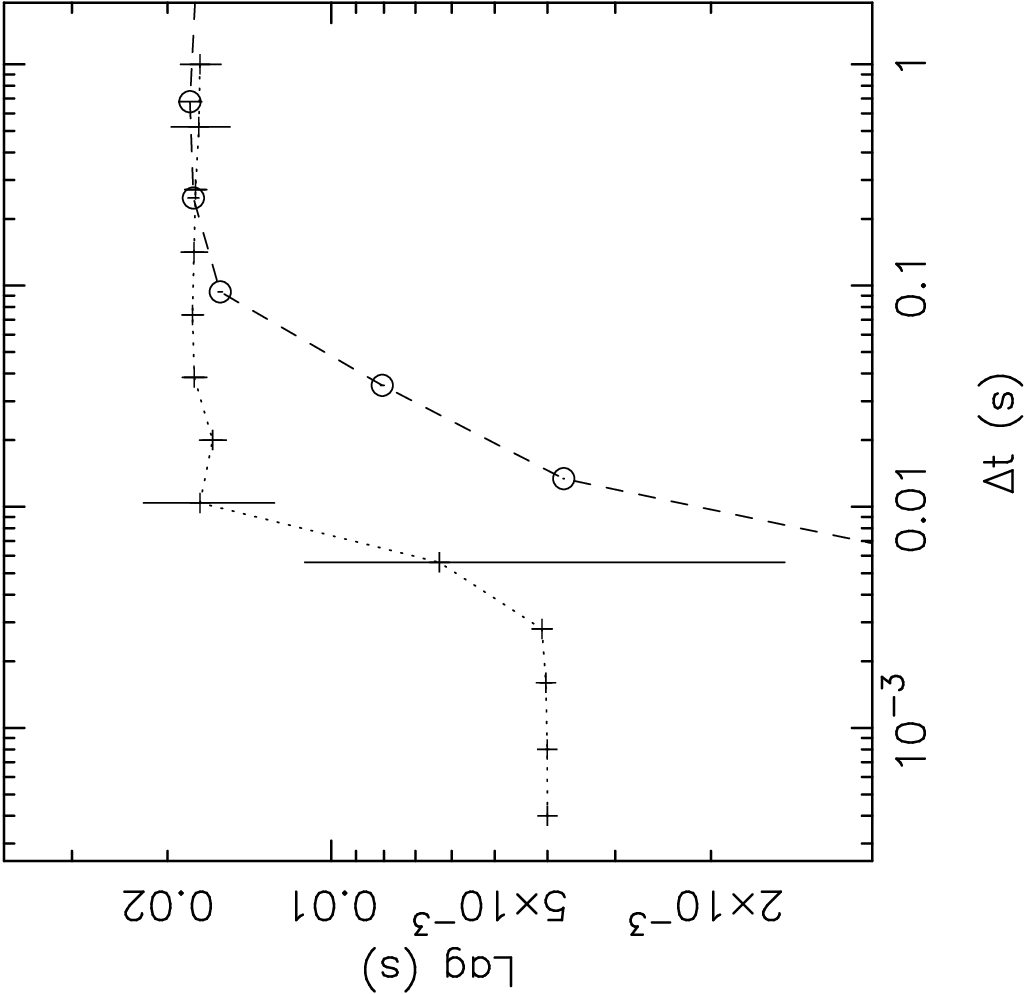}
\vspace{-7mm}
\narrowcaption{Time lags between two series of events $\{t\}$ and $\{t'\}$. 
Each series consists of three components. $\{t\}=\{x\}+\{y\}+\{z\}, 
\{t'\}=\{x'\}+\{y'\}+\{z'\}$. The first random  shot component
$\{x\}$ and $\{x'\}$ have exponentially rising and decay time $\tau=0.01$ s
and $x'(i)=x(i)+0.02$s. The second  random  shot component
$\{y\}$ and $\{y'\}$ have exponentially rising and decay time $\tau=1$ ms
and $y'(i)=y(i)+5$ms. Total shot rate is 1000 cts/s. $\{z\}$ and $\{z'\}$
are two independent white noise series with rate of 100 cts/s.   
The series duration is 2000 s. $Crosses$ are from the time domain 
technique, $Circles$ from the Fourier analysis.  
    }
\end{figure}

\clearpage
The timescale spectral method for time lag analysis is a powerful tool 
in revealing the characteristic of emission process. Applying the time domain 
technique, Qu et al. (2001) detected time lags of hard X-ray photons 
from Cyg X-1 in the short timescale region 
($\Delta t < 0.01$ s or $f > 100$ Hz) with PCA/RXTE data 
for the first time, giving strong
constraint on emission mechanism; Feng et al. (2002) revealed the
time delays of soft photons in terrestrial gamma-ray flashes (TGFs) 
observed by BATSE/CGRO, strongly supporting the discharging mechanism 
of TGF production. Investigating temporal properties of GRBs by
the timescale spectral method is in process. Figure 4 shows the spectrum 
of time lag of soft photons and the energy dependence of time lag
for GRB910503 detected by BATSE. 

\begin{figure}[ht]
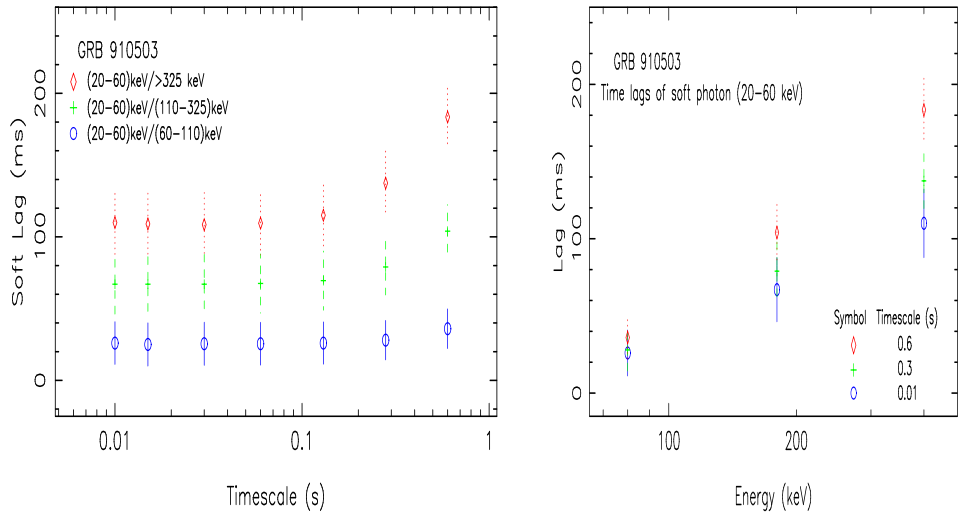

\vspace{-33mm}
\includegraphics{Fig4a.ps}
\includegraphics{Fig4b.ps}
\vspace{102mm}
\caption{Soft time lags of GRB 910503. $Left~panel$: 
Timescale spectra of time lag. $Right~panel$: Time lag of 20-60 keV photons
vs. energy of hard photons. 
    }
\end{figure}

\section{Discussion}
The complex variability of high-energy emission shown in different
time scales is a common character for X-ray binaries, super massive black
 holes and $\gamma$-ray bursts. The variability is caused by various 
physical processes at different timescales. 
For understanding the emission process of high-energy photons it is necessary
to know  the variation characteristics in different timescales quantitatively,
i.e. to derive timescale spectra from observed light curves. 
In addition to the methods of making spectral analysis for  power density and time lag
in the time domain introduced in last two sections, the algorithms to calculate  
timescale spectra for coherence, hardness, variability duration, and 
correlation coefficient between two characteristic quantities
have been also proposed (Li 2001). 

There now exist two kinds of spectral analysis: frequency
analysis and timescale analysis.
As any observable physical process always occurs in the time domain, a frequency spectrum 
obtained by frequency analysis needs to be interpreted in the time domain.
But a frequency analysis is based on a certain kind of time-frequency transformation. 
Different mathematically equivalent representations with different bases or 
functional coordinates in the frequency domain exist for certain time series,
a Fourier spectrum with the trigonometric basis does not necessarily represent the 
true distribution of a physical process in the time domain. The rms variation vs. 
timescale of a time-varying process may differ substantially from its Fourier spectrum.
Figure 1 shows that 
the Fourier transform distorts the power density distribution of a stochastic process 
at short timescales seriously. The
timescale analysis performed directly in the time domain can derive real timescale
distribution for quantities characterizing temporal property. In comparison with
the frequency analysis, timescale spectra from the timescale analysis can more exactly
represent timescale distributions of a real physical process, and more sensitively 
reveal temporal characteristics at short timescales for a stochastic process.
Until now Fourier analyses of variabilities of X-ray binaries observed by various 
instruments are all failure to detect hard X-ray lags at the high frequency region
of $f>100$ Hz. Our simulation study shows it is an intrinsic weakness of the 
Fourier method, that more sensitive X-ray detectors of next generation can still
not observe high frequency lags with Fourier technique. On the other hand, 
the timescale analysis can already derive time lag spectra at short timescales reliably
from existing data.


\begin{chapthebibliography}{1}
\bibitem{li01}
Li T.P. (2001) {\it Chin. J. Astron. Astrophys.}, 1, 313
\bibitem{lietal99}
Li T.P., Feng Y.X. \& Chen L. (1999) {\it ApJ}, 521, 789
\bibitem{limu02}
Li T.P. \& Muraki Y. (2002) {\it ApJ}, 578, 374
\bibitem{Feng}
Feng H., Li T.P., Wu M., Zha M. \& Zhu Q.Q. (2002) $GRL$, 29, No.3
\bibitem{hua}
Hua X.M., Kazanas D. \& Titarchuk L. (1997) $ApJ$, 482, L57
\bibitem{ka}
Kazanas D., Hua X.M. \& Titarchuk L. (1997) $ApJ$, 480, 735
\bibitem{No}
Nowak M.A., Wilms J., Vaughan B.A., Dove J.B. \& Begelman C.  (1999) $ApJ$, 510, 874
\bibitem{Qu}
Qu J.L. \& Li T.P. (2001) {\it Acta Astron. Sin.}, 42, 140

\end{chapthebibliography}
\end{document}